\documentclass[preprint,12pt]{elsarticle}
\usepackage{graphicx}
\usepackage{amssymb} 
\usepackage{epstopdf}
\usepackage{listings}
\usepackage{float}
\usepackage{lineno}
\usepackage{etex}
\usepackage{graphicx,color,colordvi}
\usepackage{latexsym,amsbsy,epsfig,fancybox}
\usepackage{amstext}
\usepackage{subfigure}
\usepackage{amsfonts,textgreek}
\usepackage{mathrsfs}
\usepackage{mathtools}
\usepackage{stmaryrd,dsfont}
\usepackage{bbm}
\usepackage{url}
\usepackage{wrapfig,bm}
\usepackage{tikz}
\usepackage{pstricks}
\usepackage{caption}
\usepackage{psfrag}
\usepackage{upgreek}
\usepackage{isomath}
\usepackage{latexsym}
\usepackage{array}
\usepackage{multirow}
\usepackage{booktabs}
\usepackage{verbatim}
\usepackage{natbib}
\biboptions{sort&compress}
\usepackage{color}
\usepackage{colortbl}
\usepackage[colorlinks=true,citecolor=blue]{hyperref}
\usepackage{overpic}
\usepackage{enumerate}
\usepackage{amsmath,amssymb,xspace}
\usetikzlibrary{plotmarks}
\usetikzlibrary{fit,positioning,arrows}
\usetikzlibrary{shapes.geometric, plotmarks}
\usepackage{appendix}

\usepackage[margin=3cm]{geometry}
\usepackage{soul}

\newcommand{\ac}[1]{\textcolor{red}{add citation}}

\tikzstyle{nicebox}=[draw=black!100, fill=white!10, rectangle, inner sep=4pt, inner ysep=16pt]
\tikzstyle{niceboxtitle}=[draw=black!100, fill=white, text=black, rectangle]

\usetikzlibrary{arrows,decorations.markings}

\journal{arXiv}
\begin{document}
\begin{frontmatter}

\title{Network evolution controlling strain-induced damage and self-healing of elastomers with dynamic bonds}

\author[1]{Yikai Yin\fnref{label1}}
\address[1]{
 Department of Materials Science and Engineering, Stanford University, Stanford, CA 94305
}
\author[2]{Shaswat Mohanty\fnref{label1}}
\address[2]{
 Department of Mechanical Engineering, Stanford University, Stanford, CA 94305
}
\author[3]{Christopher B. Cooper}
\author[3]{Zhenan Bao}
\address[3]{
 Department of Chemical Engineering, Stanford University, Stanford, CA 94305
}
\author[2]{Wei Cai}
\fntext[label1]{Equal contribution}


\begin{abstract}
Highly stretchable and self-healable supramolecular elastomers are promising materials for future soft electronics, biomimetic systems, and smart textiles, due to their dynamic cross-linking bonds. 
The dynamic or reversible nature of the cross-links gives rise to interesting macroscopic responses in these materials such as self-healing and rapid stress-relaxation. However, the relationship between bond activity and macroscopic mechanical response, and the self-healing properties of these dynamic polymer networks (DPNs) remains poorly understood.
Using coarse-grained molecular dynamics (CGMD) simulations, we reveal a fundamental connection between the macroscopic behaviors of DPNs and the shortest paths between distant nodes in the polymer network. Notably, the trajectories of the material on the shortest path-strain map provide key insights into understanding the stress-strain hysteresis, anisotropy, stress relaxation, and self-healing of DPNs. Based on CGMD simulations under various loading histories, we formulate a set of empirical rules that dictate how the shortest path interacts with stress and strain. This lays the foundation for the development of a physics-based theory centered around the non-local microstructural feature of shortest paths to predict the mechanical behavior of DPNs.
\end{abstract}

\end{frontmatter}%


\section{Introduction}

Supramolecular elastomers with high stretchability and self-healing ability are desirable components in a wide range of engineering applications, including flexible electronics~\cite{oh2019stretchable,kang2019self,cooper2023autonomous}, biomaterials~\cite{brochu2011self} and energy storage~\cite{mai2020self}. These elastomers incorporate dynamic bonds, e.g., hydrogen bonds and metal-ligand coordination bonds, as cross-links that can easily break and re-form between polymer chains~\cite{appel2012supramolecular,cooper2022using}. Bond breaking events enhance the stretchability of elastomers, but also introduce stress-strain hysteresis under cyclic loading, indicating strain-induced damage. Once the external load is released, bond re-forming events can progressively restore the original property during a long-time relaxation, resulting in self-healing. While the key role of the bond-breaking and re-forming events was clearly identified in experiments~\cite{cordier2008self,li2016highly}, their connection to the mechanical property remains unclear. For example, given a certain type and density of dynamic bonds of an elastomer, the questions of how the damage evolves with strain and how fast and to what extent the elastomer can be healed over time remain open~\cite{yu2018mechanics,stukalin2013self}. Therefore, predicting the stretchability and self-healing efficiency of supramolecular elastomers prior to synthesis is still a major challenge~\cite{burnworth2011optically}. 

A key to addressing this challenge is to obtain an explicit bond-property connection for these dynamic polymer networks (DPNs). We believe this connection lies in the microstructure, specifically the network topology of the DPNs, which evolves during deformation and self-healing.
Probing the microstructural mechanisms of strain-induced damage and self-healing in DPNs through experimental techniques have been suggested in \citep{lloyd2023covalent}, however, they have not been implemented extensively.
%
Here we employ coarse-grained molecular dynamics (CGMD) simulations to study the deformation and self-healing mechanisms of DPNs in terms of their microstructure. Our CGMD simulations reveal how the bond-breaking and re-forming events are organized in the bulk during straining and healing, in contrast to the existing CGMD works which primarily focus on the interfacial mechanisms~\cite{stukalin2013self,ge2014healing}. Our CGMD model successfully captures the (i) stress-strain hysteresis during a loading-unloading cycle; (ii) strain-induced anisotropy after the loading cycle; and (iii) self-healing during a long relaxation process at the unloaded state.
We analyze the topology of our CGMD model and show that the shortest paths between distant nodes in the polymer network are fundamentally connected to the macroscopically exhibited behavior of DPNs. Notably, the evolution of the shortest path-strain map provides key insights into understanding the stress-strain hysteresis, anisotropy, stress relaxation, and self-healing of DPNs. Based on the observations from a series of CGMD simulations (numerical experiments), we formulate a set of empirical rules that dictate how the shortest path interacts with stress and strain, and how it consequently explains the macroscopic behavior exhibited by these materials. This analysis serves as the starting point for the development of a physics-based theory for the mechanical behavior of DPNs.
Our findings offer valuable insights into the behavior of existing self-healing elastomers, and can potentially guide the design process for future DPN materials.

The rest of the paper is structured as follows.  Section~\ref{sec:cgmd_model} describes our CGMD model of DPN. Section~\ref{sec:results} presents the CGMD simulation results, showing that the model successfully captures the stress-strain response and self-healing behavior observed experimentally. 
Section~\ref{sec:discussion} shows how the shortest path analysis of the polymer network reveals insights into the fundamental mechanisms of strain-induced damage and self-healing. 
Section~\ref{sec:sp_laws} presents a set of empirical rules regarding how shortest paths in DPN are connected to stress and strain, summarized from CGMD simulations under different loading histories. Section~\ref{sec:conclusions} concludes the article and provides an outlook on the pathway towards formulating a physics-based analytic theory on the mechanical properties of DPNs.

\section{CGMD Model} \label{sec:cgmd_model}
We performed CGMD simulations using LAMMPS~\cite{plimpton1995fast} on a well-established bead-spring model~\cite{kremer1990dynamics} subjected to periodic boundary conditions. Each model contains 500 chains, each with 500 beads. Each bead represents a group of atoms and is assumed to interact with other beads through empirical potential functions and follows Newton's equations of motion. The CGMD simulation cell is subjected to periodic boundary conditions in all three directions. Non-bonded interactions between each pair of beads were modeled by a Lennard-Jones (LJ) potential. The backbone interactions between neighboring beads on the same chain were modeled by the finite extensible nonlinear elastic (FENE) potential~\cite{kremer1990dynamics}. The FENE bonds are unbreakable, meaning that we assume the backbone bonds (e.g., covalent bonds) are much stronger than the dynamic bonds (e.g., hydrogen bonds). After equilibration of the polymer melt with a two-step protocol~\cite{sliozberg2012fast}, we introduced two extra types of beads in addition to the rest of the backbone beads (which are labeled as Type-C). Type-A beads are stickers (representing hydrogen-bonding units, for example, isophorone bisurea or hexamethylene bisurea) equally spaced on the backbone chains. Type-B beads are cross-linking agents initially floating in space. The ratio of type-A beads to type-B beads is 2:1.
This type of structure is motivated by \cite{cooper2020multivalent}, where
each cross-linking bond forms in an A-B-A motif where one type-B bead connects two type-A beads, as demonstrated in Fig.~\ref{fig:fig-stress-strain-1}(a). We used a strong LJ potential to model the dynamic interaction between beads A and B. When beads A and B are close enough, they attract each other and form a bond. When beads A and B are pulled apart further than a cut-off distance, their interactions drop to zero and the bond is considered broken. 
The parameters of the LJ potential for the A-B-A interaction are chosen so that one B-type bead can only bond to two A-type beads, so that an agglomeration of A-type (or B-type) beads is not energetically favorable. This is achieved by (i) modeling a significantly stronger LJ potential for the A-B pair and (ii) by modeling an exclusively repulsive interaction between the A-A and B-B bonds by setting the cut-off as $2^{1/6}\sigma_{\rm i-i}$, where ${\rm i-i}$ refers to the interaction between atoms of type ${\rm i}$, which in this case corresponds to type A and B beads. The other pairwise interactions have a higher cut-off to account for both repulsive and attractive interactions.
After equilibration, we obtain a DPN configuration with around 8000 A-B-A cross-links. The resulting structure is in a cubic simulation cell of length 98.3~nm. More details on the model are given in the Methods. The deformation of the modeled elastomer is assumed to be volume-preserving or incompressible, owing to the nearly incompressible behavior of most elastomers\cite{melly2021review}.

\section{Results} \label{sec:results}
\subsection{Stress-strain hysteresis}
\begin{figure}[H]
\centering
\includegraphics[width=1\linewidth] {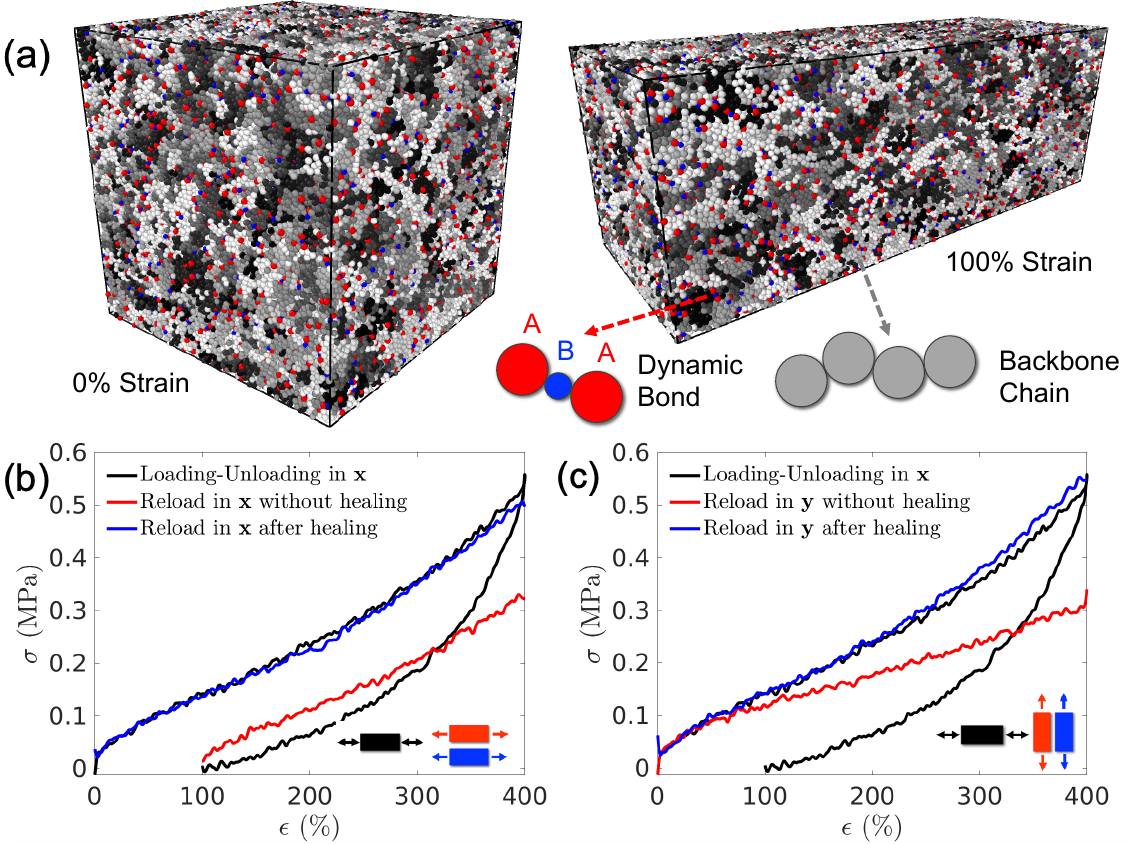}
\caption{(a) Snapshots of an elastomer with dynamic cross-linking bonds extracted by CGMD. (b) and (c) Stress-strain curves during loading-unloading in the $x$-direction, reloading with and without healing in the $x$-direction and $y$-direction.
} 
\label{fig:fig-stress-strain-1}
\end{figure}

The equilibrated DPN model is loaded in uniaxial tension along the $x$-direction at a strain rate of $8.7\times 10^5 \, {\rm s}^{-1}$ until $400\%$ strain, followed by unloading (i.e. reversing the sign of strain rate) until zero stress is reached. The volume is conserved during the loading-unloading process.  The time period of loading from $0$ to $400\%$ strain is $~\sim 4.6 \, \mu{\rm s}$.
The residual strain at the end of unloading is about $100\%$. As shown in Fig.~\ref{fig:fig-stress-strain-1}(b), the stress-strain curves during loading and unloading exhibit a hysteresis which is commonly referred to as the Mullins effect~\cite{mullins1965stress}. Although the Mullins effect is commonly applied to the behavior exhibited by filled polymers (polymer networks with filer particles, the phenomenon is common to all polymers, including polymers without fillers.
%
The stress-strain curve during unloading is lower than that during loading, indicating that certain bond-breaking events have happened during loading that weakened the polymer network.

We made three copies of the configuration obtained at the end of the loading-unloading cycle and subjected them to different loading paths.
The first configuration is immediately loaded in tension along the $x$-direction again.  The resulting stress-strain curve is shown in Fig.~\ref{fig:fig-stress-strain-1}(b) (red line).
The second configuration is immediately loaded in tension along the $y$-direction. The resulting stress-strain curve is shown in Fig.~\ref{fig:fig-stress-strain-1}(c) (red line).
The differences in these two stress-strain curves show that the material becomes anisotropic after the first loading-unloading cycle.
It is also interesting to note that the stress-strain curve along the second loading in the $y$-direction is lower than the initial stress-strain curve.  This means that the first loading-unloading cycle along $x$ also weakens the material in the $y$-direction.

\subsection{Self-healing behavior} 

The third configuration is allowed to relax under zero stress for a long period of $107.4\,\mu$s.
Remarkably, during this period the strain in the $x$-direction is seen to go back to close to zero ($\sim 2\%$ strain), and the simulation cell eventually returns close to its initial cubic shape.
Furthermore, when the resulting configuration is subjected to uniaxial loading again in $x$ and $y$ directions, the stress-strain curves recover the original stress-strain curve, as shown in Fig.~\ref{fig:fig-stress-strain-1}(b) and (c), respectively (blue lines).
This means that our CGMD model successfully captures the self-healing behavior of DPNs.
After the self-healing period, the elastomer not only returns close to its initial shape but also regains its original mechanical properties.
In this work by self-healing we mean the process of DPN recovering its original shape and stress-strain properties after the strain-induced damage by the first loading-unloading cycle, not the process of rejoining two halves of the elastomer after being cut by a blade.

\begin{figure}[H]
\centering
\includegraphics[width=1\linewidth] {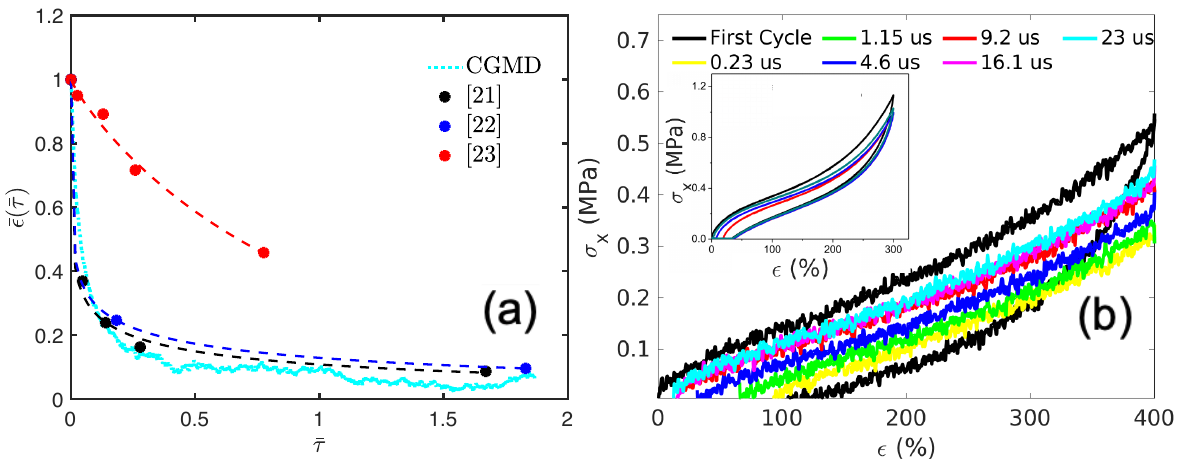}
\caption{{(a) Recovered strain fraction, $\bar{\epsilon}(\bar{\tau})={\epsilon(\bar{\tau})}/{\epsilon_s}$, from the CGMD simulation compared against experimental literature of dynamically cross-link networks~\cite{wang2016dynamic,peng2018strong,li2022robust}.} (b) Stress-strain curve for reloading at the end of different extent of healing time, $\tau_H$ with $\tau \approx 62\,\mu$s (inset from \cite{peng2018strong}, where $\tau\approx 5.5\, {\rm min}$; for the inset: black: original, \textcolor{red}{red}: $\tau_H=0\, {\rm min}$, \textcolor{green}{green}: $\tau_H=1\, {\rm min}$, \textcolor{blue}{blue}: $\tau_H=10\, {\rm min}$).}
\label{fig:fig-stress-heal-times}
\end{figure}

To make a qualitative comparison of the kinetics of the self-healing process with experiments, we plot the fraction residual strain $\bar{\epsilon} = \epsilon/\epsilon_{\rm s}$ as a function of dimensionless time $\bar{\tau} = t / \tau$, where $\epsilon_{\rm s}$ is the residual strain at the beginning of the self-healing period, and $\tau$ is a characteristic time obtained by fitting the strain $\epsilon(t)$ to an exponential function ($\epsilon_{\rm s}\exp(-t/\tau)$).
Although due to computational limitations, the absolute time scale of self-healing in the CGMD simulations is much faster than that in the experiments, the non-dimensionalized relaxation curve $\bar{\epsilon}(\bar{\tau})$ predicted by CGMD shows qualitative agreement with some of the experiments, as shown in Fig.~\ref{fig:fig-stress-heal-times}(a).
In particular, the relaxation is characterized by a rapid period where a large fraction (e.g. 60\%) of the strain is recovered over a short period of time (e.g. $\bar{\tau} \in [0, 0.1]$), followed by a much longer period where the remaining strain is recovered.
We also extract the configurations from the self-healing process after it has proceeded for different amounts of time, and subject these configurations to uniaxial tensile loading in $x$-direction again.
Fig.~\ref{fig:fig-stress-heal-times}(b) plots the corresponding stress-strain curves; they are qualitatively similar to the experimental results~\cite{wang2016dynamic,peng2018strong,li2022robust} shown in the inset.
In particular, the initial slope and the general shape of this family of stress-strain curves are similar to each other.  The longer the healing time, the higher is the peak stress reached at re-loading (at the final strain of $400\%$).

\section{Discussion} \label{sec:discussion}

\subsection{Non-local network characterstics} \label{sec:nonlocal}
We have seen that the CGMD model with reversible (A-B-A type) cross-links is able to capture both the stress-strain hysteresis and the self-healing behavior of DPNs.
This CGMD model thus provides us with an opportunity to understand the physical origin of these behaviors.  We start with three general remarks on how we may approach this problem.
First, we can establish that these behaviors are caused by the bond-breaking and re-forming events that alter the network topology, because these behaviors disappear when we change the cross-links to be permanent (i.e. unbreakable) resulting in the absence of bond-breaking events in the CGMD model~\cite{yin2020topological}.

Second, many existing theories that link the stress-strain responses of polymer networks to bond-breaking events are based on the length of \emph{local chains} between two neighboring cross-link sites on the same backbone.
For example, the network alteration theory~\cite{chagnon2006development,marckmann2002theory} attributes the changes in the stress-strain curve to the changes in the distribution of local chain lengths, with more recent works extending this analysis to polymers with dynamic bonds~\cite{yu2018mechanics,vernerey2018transient}.
However, it can be easily shown that the local chain length cannot possibly explain the behaviors reported here.
In our CGMD simulations, most of the reversible bonds immediately form (with other partners) after breaking, so that at any instant, $>99\%$ of all Type-A (and Type-B) beads remain bonded.
Since the Type-A beads are regularly distributed on the polymer backbone (with a fixed period of 15 beads), the local chain length remains very close to 15 during the entire process of loading, unloading, and self-healing.
Therefore, it can be concluded that the rich behaviors exhibited by DPNs are not related to the local chain length (which remains at 15) but are the result of the changing network topology.  In other words, although every Type-A bead stays connected (through the A-B-A cross-link) to another Type-A bead nearly all the time, what is important is which ones are connected to each other.
This connectivity information can be specified by a \emph{connectivity matrix} $\mathbf{A}$, which is an $n\times n$ matrix, with $n$ being the number of Type-A beads; $A_{ij} = 1$ if bead $i$ and bead $j$ are cross-linked and $A_{ij} = 0$ if they are not.
We can thus envision the scenario that during deformation (or healing), the $\mathbf{A}$ matrix evolves with strain (or time), and hence changes the stress-strain characteristics of the DPN.
While we believe that the connectivity matrix formally captures the network characteristics responsible for the stress-strain hysteresis and self-healing behaviors, it is a microscopic description of the network that still contains many irrelevant details.
For example, another realization of the CGMD model of the DPN with the same general procedure but different initial configurations (e.g. using different random seeds) would result in a different connectivity matrix but a DPN model with nearly identical deformation and self-healing behaviors.
Therefore, we still need to pinpoint the essential microstructural features contained in the connectivity matrix that are responsible for the deformation and self-healing behaviors of DPNs.

Third, a recent study has shown that the distribution of shortest-path lengths between far-away beads in the polymer network is the controlling microstructural parameter responsible for the stress-strain hysteresis of elastomer with breakable and irreversible cross-links~\cite{yin2020topological}.
We note that the shortest-path (SP) lengths between far-away beads are a non-local property of the polymer network; hence an SP-based theory will not suffer from the same limitation as the theories based on local chain lengths mentioned above.
Therefore, it is of interest to see if SP lengths also control the deformation and self-healing behaviors of DPNs reported here.
However, the reversible bonds in DPNs make the situation much more complex than traditional elastomers with irreversible bonds.
In the language of the connectivity matrix, all that can happen for an elastomer with irreversible bonds is that certain matrix element $A_{ij}$ can change from $1$ to $0$ (and stays $0$ afterwards).
However, in DPN a matrix element $A_{ij}$ can also change from $0$ to $1$. In fact, for the choice of the interaction potential in our DPN model, the total number of non-zero elements stays nearly the same all the time.

\subsection{Shortest path analysis}

\begin{figure}[H]
\centering
\includegraphics[width=0.75\linewidth] {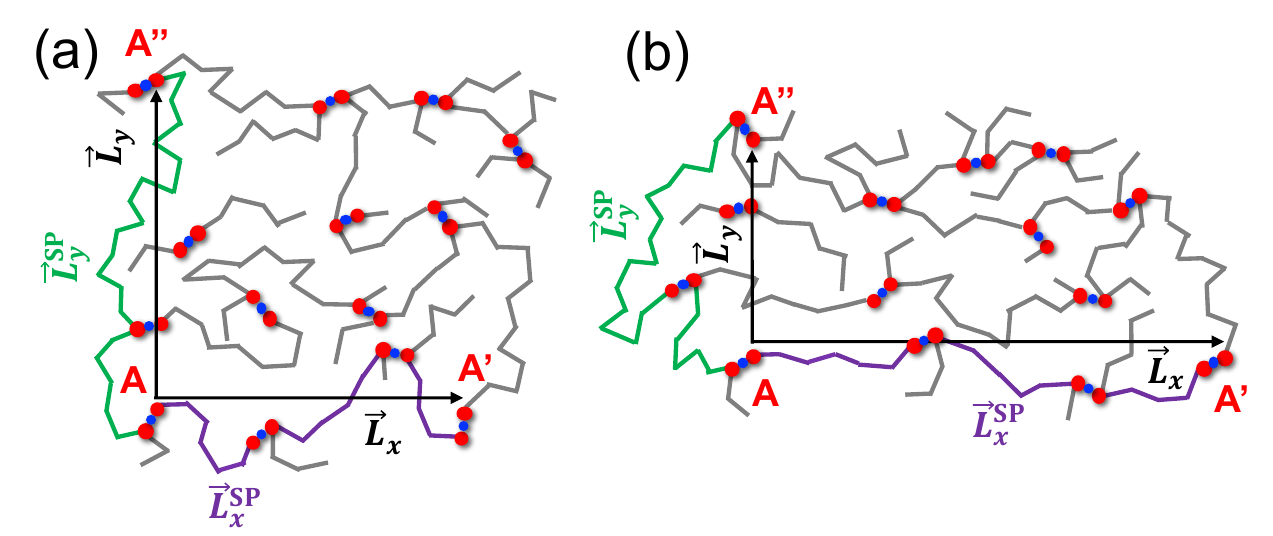}
\caption{Schematic of the polymer chain network (a) before and (b) after loading in the $x$-direction, where A-B-A bonds are marked with red-blue-red dots and backbone bonds are marked with grey line segments. A-A' and A-A'' are example pairs of vertices separated by vectors $\vec{L}_x$ and $\vec{L}_y$. The shortest paths (SPs) connecting A-A' and A-A'' are colored purple and green, respectively.} 
\label{fig:fig-sp-sche}
\end{figure}

To reduce the computational cost without the loss of information about the network connectivity,
we define a graph that contains only Type-A beads as vertices and weighted edges between them.
Two vertices that are connected by a cross-link result in an edge with weight $1$.
Two vertices that are connected by a polymer backbone result in an edge with a weight equal to the length of the polymer chain between them (which is always $15$ in our model).
We then choose an offset vector $\vec{q}$.
For each vertex (e.g. A), we identify the vertex (e.g. A') that is nearest to the location offset from A by $\vec{q}$ and then find the shortest path (SP), which is the contour length traveled over all the bonds between them.
This shortest path search is performed for all vertices for the same offset vector $\vec{q}$~\cite{yin2020topological}.
When the polymer network is deformed, the offset vector $\vec{q}$ is subjected to the affine transformation consistent with the macroscopic strain.
In this work, we focus on the cases where $\vec{q}$ is a repeat vector of the simulation box, $\vec{L}_x$ in the $x$-direction or $\vec{L}_y$ in the $y$-direction, as illustrated in Fig.~\ref{fig:fig-sp-sche}. For this specific choice of $\vec{q}$, vertex A' will always be the periodic image of vertex A in $x$-direction or $y$-direction, respectively.
According to this definition, the shortest paths would not change with deformation if there were no bond-breaking or bond-forming events that would cause a change in the network connectivity.

\begin{figure}[H]
\centering
\includegraphics[width=0.75\linewidth] {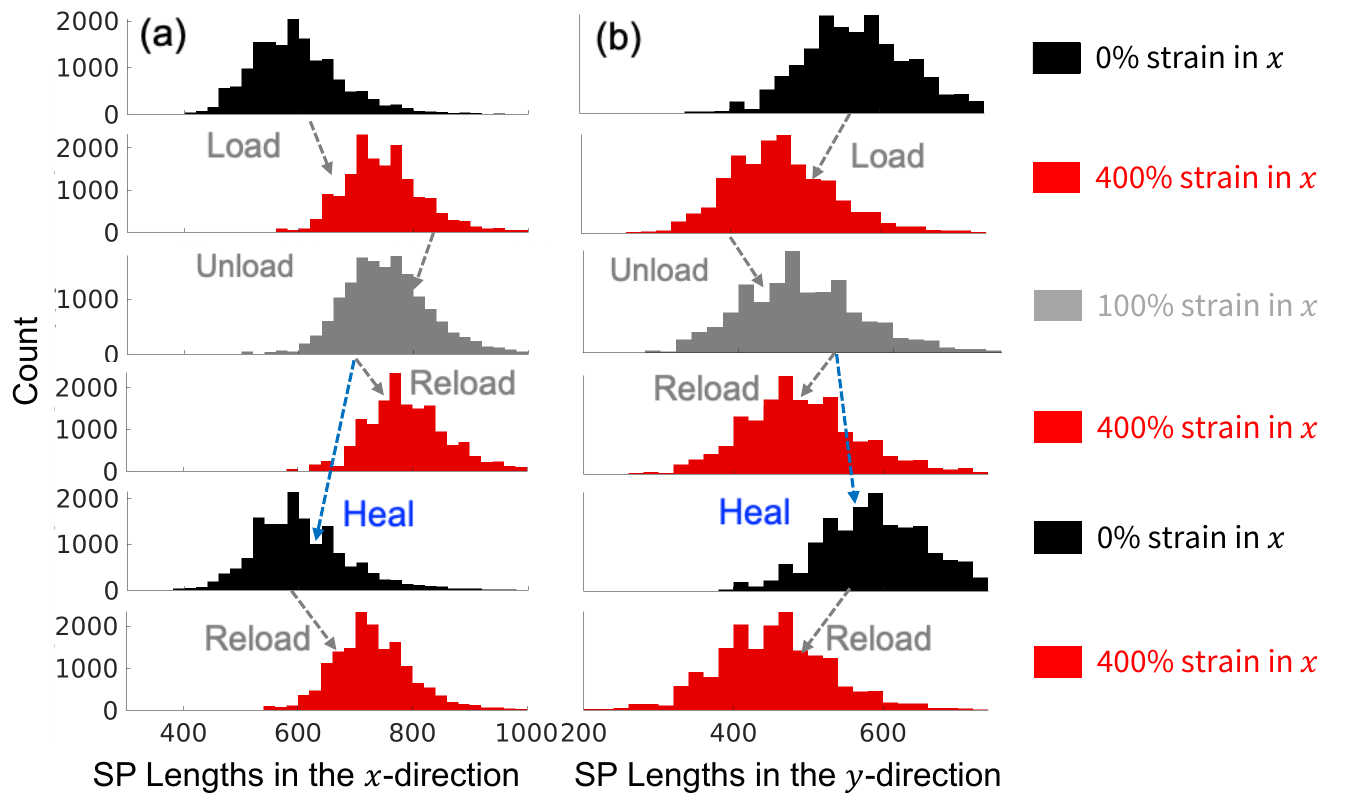}
\caption{Distribution of shortest path (SP) lengths (a) in the $x$-direction and (b) $y$-direction with strain in the $x$-direction. The colors of the histogram denote the state of strain. The text next to the arrow refers to the deformation process that transforms one configuration into another. 
}
\label{fig:fig-sp-hist-load}
\end{figure}

Fig.~\ref{fig:fig-sp-hist-load}(a) shows the evolution of the SP length distribution of beads offset by $\vec{q} = \vec{L}_x$ during the deformation shown in Fig.~\ref{fig:fig-stress-strain-1}(b). Before initial loading, the SP length follows a Gaussian-like distribution, reflecting the randomness of the initial polymer structure. During the initial loading, this histogram noticeably shifts to the right (towards longer SPs) due to a large number of bond-breaking events, which destroy the original SPs, and replace them with longer SPs. 
The SP distribution changes negligibly during unloading.
When the polymer is re-loaded in $x$ immediately after unloading, the SP length distribution also stays nearly unchanged.
That the SP length distribution (for $\vec{q}$ along the loading direction) increases during loading and remains nearly unchanged during unloading and immediate reloading is similar to the observation in polymer networks with irreversible bonds~\citep{yin2020topological}.

Figure~\ref{fig:fig-sp-hist-load}(b) shows that loading in $x$ causes the SP length distribution in $y$ to shift to the left (for $\vec{q} = \vec{L}_y$), i.e., the SPs in $y$ become shorter. This is different from polymer networks with irreversible bonds where the SP lengths in the transverse direction remain nearly unchanged during loading~\cite{yin2020topological}. We attribute this unique behavior of the anisotropy in the SP evolution observed in the DPNs to the reversible bonds, which are able to create shorter SPs in $y$ while the cell size in $y$ decreases with strain in $x$. 
%

Remarkably, the SP length distributions in both $x$ and $y$ directions return to the initial pattern (i.e., a pronounced shift to the left) during the healing process.
During subsequent reloading, the SP length distributions change with strain in the same way as during the initial loading.
This indicates that the damage and anisotropy caused by the initial loading are eliminated by the healing process.
The strong correlation between SP lengths and stress-strain hysteresis during loading, unloading, and self-healing leads us to conclude that the SP is a key microstructural feature controlling the mechanical properties of DPNs.
%
%
\subsection{Shortest path evolution under different loading paths}

\begin{figure}[H]
\centering
\includegraphics[width=1\linewidth] {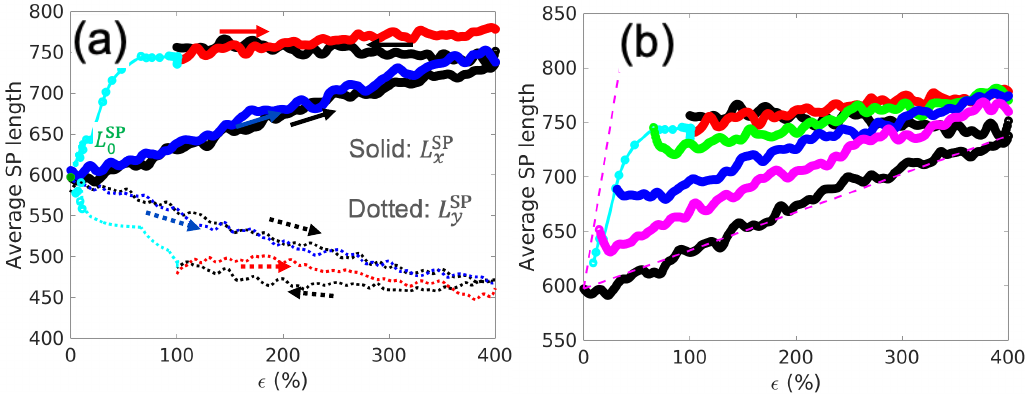}
\caption{(a) Average SP length ($L_x^{\rm SP}$ and $L_y^{\rm SP}$) as a function of strain in the $x$-direction during the loading histories (black: loading-unloading, \textcolor{red}{red}: reloading instantaneously, \textcolor{blue}{blue}: reloading after healing, \textcolor{cyan}{cyan}: healing). 
(b) Average SP ($L_x^{\rm SP}$)  
for loading cycles at different extents of $\tau_H$ (black: loading-unloading, \textcolor{red}{red}: reloading instantaneously, \textcolor{green}{green}: $\tau_H=1.15\, \mu{\rm s}$, \textcolor{blue}{blue}: $\tau_H=4.6\, \mu{\rm s}$, \textcolor{magenta}{magenta}: $\tau_H=16.1\, \mu{\rm s}$, dashed magenta lines indicate the bounds of the SP evolution).
}
\label{fig:fig-avg-sp-load}
\end{figure}

The stress-strain behavior and bond dynamics of DPN are more complicated than of polymer networks with irreversible bonds.  For comparison purposes, let us first give a quick summary of the situation for polymer networks with irreversible bonds.  There, the stress-strain curve remains reversible and SP distribution does not change during unloading and reloading as long as the previous maximum strain is not exceeded, because no new bonds break and no new bonds can form~\cite{ducrot2014toughening}.  Once the previous maximum strain is exceeded, bond-breaking events occur resulting in an irreversible softening of the material as is the characteristic of the Mullins effect~\cite{yin2020topological}. Hence the entire stress-strain behavior for loading/unloading along a given direction can be described by a family of curves parameterized by the maximum strain. This relatively simple picture no longer applies to DPNs due to their ability to form new bonds.  To probe the behavior of DPNs, we performed CGMD simulations and the SP analysis along different loading/unloading paths.

Firstly, we analyze the shortest path evolution during the loading trajectory shown in Fig.~\ref{fig:fig-stress-strain-1}(b-c). 
Fig.~\ref{fig:fig-avg-sp-load}(a) shows the average SP (in both $x$ and $y$ directions) as a function of strain during loading to $400\%$, unloading, healing and reloading.
The SP-strain map shown here provides a more concise summary of the SP evolution shown in Fig.~\ref{fig:fig-sp-hist-load}.
It is interesting to note that on the SP-strain map the healing process can be clearly seen to occur in two steps.  The first step corresponds to a (rapid) reduction of strain with nearly no change of SP in $x$, and the second step corresponds to a (slow) reduction of both SP in $x$ and strain.
The first, rapid, step of healing can also be seen in Fig.~\ref{fig:fig-stress-heal-times}.
Fig.~\ref{fig:fig-avg-sp-load}(b) shows the average SP in $x$-direction during loading after the DPN has been allowed to heal for different amounts of time $\tau_H$.
Interestingly, it appears that they can be well approximated by a family of linear curves with different initial points but a common final point (at $400\%$ strain).

\begin{figure}[H]
\centering
\includegraphics[width=1\linewidth] {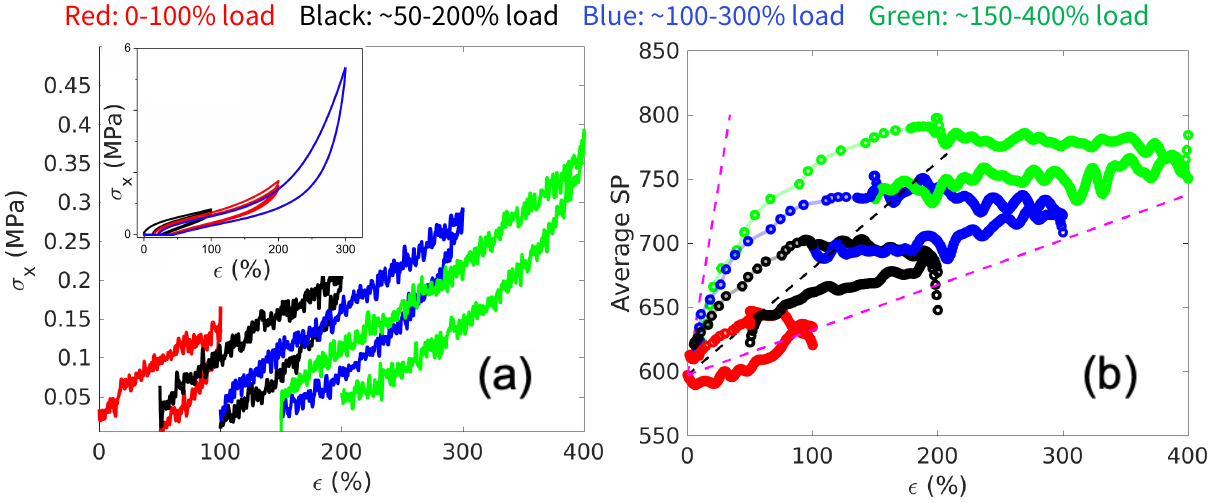}
\caption{
{(a) Tensile stress during loading-unloading in $x$ measured in the CGMD simulation (inset: experimental results from \cite{peng2018strong}. (b) Average SP length ($L_x^{\rm SP}$) as a function of strain in the $x$-direction (dashed magenta lines indicate the bounds of the SP evolution; dashed black lines indicate the end of point of unloading where the stress goes to zero). }
} 
\label{fig:multi_cycle}
\end{figure}

Secondly, we consider a loading path where the sample is repeatedly unloaded to zero stress and then re-loaded to a higher strain.
Fig.~\ref{fig:multi_cycle}(a) shows the stress-strain curves where significant hysteresis can be observed, similar to the experimental observation (inset).
Fig.~\ref{fig:multi_cycle}(b) shows the corresponding map of SP (in $x$) versus strain.
It can be seen that during immediate reloading the SP appears more or less constant until the previous maximum strain is reached, beyond which the SP increases with strain at almost the same rate as during the initial loading.
Also shown in Fig.~\ref{fig:multi_cycle}(b) are SP-strain curves during healing from different unloaded states.
%

\begin{figure}[H]
\centering
\includegraphics[width=1\linewidth] {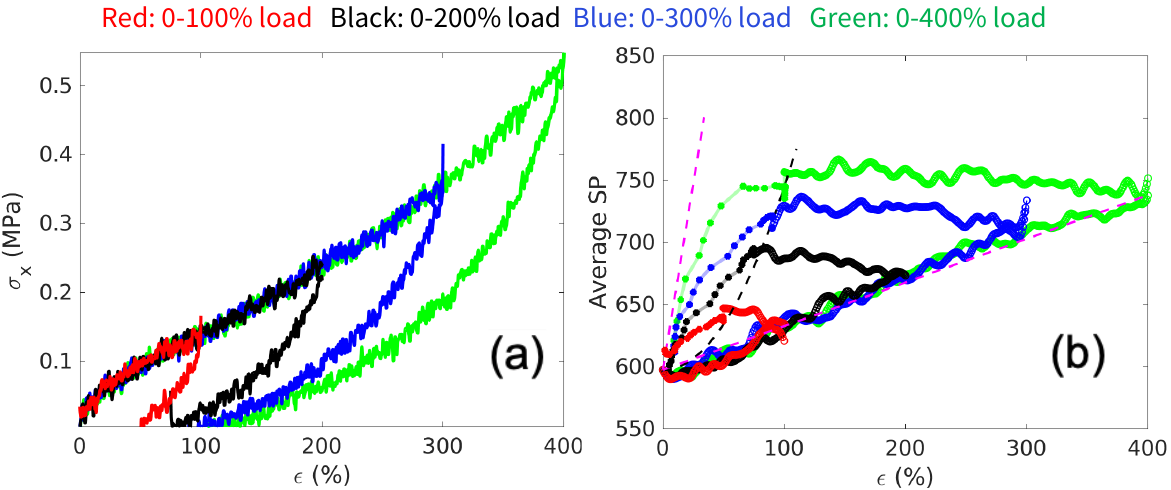}
\caption{
{(a) Tensile stress during loading-unloading in $x$ measured in the CGMD simulation. (b) Average SP length ($L_x^{\rm SP}$) as a function of strain in the $x$-direction and its subsequent evolution during healing (dashed magenta lines indicate the bounds of the SP evolution; dashed black lines indicate the point during unloading where the stress goes to zero). }}
\label{fig:slack_cycle}
\end{figure}

Finally, we consider loading paths where the sample is stretched to different extents of maximum strain, followed by unloading and healing.
Fig.~\ref{fig:slack_cycle}(a) shows the stress-strain curves and Fig.~\ref{fig:slack_cycle}(b) shows the corresponding map of SP (in $x$) versus strain.
%

\subsection{General rules of shortest path evolution}
\label{sec:sp_laws}

Although the evolution of SP depends on the loading path in a complex way, there are some general rules that can be stated based on the observations from the CGMD simulations. These empirical rules can serve as a basis for the future development of a comprehensive theory on the microstructural evolution of DPNs.
For simplicity, we limit the scope of our discussion to uniaxial tensile loading along $x$-direction while keeping the volume constant.

\paragraph{Rule 1: Bounds on average SP}
The initial average SP, $L_0^{\rm SP}$, after equilibration and before any deformation, is a function of the cross-link density (which is fixed in this study). The subsequent evolution of the average SP with strain seems to stay within a lower bound and upper bound for all the loading paths considered in this study.
\begin{equation}
    1 + \alpha \, \epsilon_x \le L_{x}^{\rm SP} / L_{0}^{\rm SP} 
    <  1 + \epsilon_x.
\end{equation}
The lower and upper bounds are shown by the magenta dashed lines on the Figs.~\ref{fig:fig-avg-sp-load}(b),~\ref{fig:multi_cycle}(b) and \ref{fig:slack_cycle}(b).
The lower bound of the average SP is reached during the initial loading, during which the average SP increases linearly with strain.  Subsequent unloading moves the average SP away from the lower bound.
We expect that the parameter $\alpha$ that characterizes the lower bound to depend on the strain rate (lower $\alpha$ at a higher strain rate), although we have not varied the strain rate in this study.

The upper bound of the average SP is obtained by considering the idealized scenario where the DPN is at full equilibrium under a non-zero strain $\epsilon_x$.
Hypothetically, this equilibrium state can be reached by first removing all the cross-links, i.e. returning the DPN to a polymer melt, deforming the melt to strain $\epsilon_x$, and re-inserting the cross-links.
This hypothetical equilibrium state represents the lowest free-energy state at any strain.  Hence there is a thermodynamic driving force for the DPN to evolve with time towards this limit.  However, there seems to be high kinetic barrier that prevents this limit from ever being reached at any $\epsilon_x > 0$. During self-healing the evolution of the SP-strain diagram appears to asymptotically approach this limit, as shown in Figs.~\ref{fig:fig-avg-sp-load}(b),~\ref{fig:multi_cycle}(b) and \ref{fig:slack_cycle}(b).

\paragraph{Rule 2: Rate of average SP increase during loading} For well-equilibrated samples,
\begin{align}
    \dot{L}_{x}^{\rm SP}  &= \beta \, L_{0}^{\rm SP}
    \quad {\rm for} \, \sigma_x > \sigma_{\rm c} 
    \ {\rm and} \ \dot{\epsilon}_x > 0 
\end{align}
The rate parameter $\beta$ seems to be a constant during the initial loading of well-equilibrated samples and does not appear to depend on the magnitude of stress as long as it is above some threshold.  The threshold stress $\sigma_c$ is much lower than the applied stress during loading.
During unloading ($\dot{\epsilon} < 0$) and immediate re-loading the average SP can be seen to still increase somewhat, but at a slower rate.
However, if the DPN is left to self-heal for some time (not necessarily all the way to the initial strain), then upon re-loading the SP increases at the same rate parameter $\beta$ as before, as shown in Fig.~\ref{fig:fig-avg-sp-load}(b).

\paragraph{Rule 3: Rate of average SP decrease during healing}
\begin{align}
    \dot{L}_{x}^{\rm SP} &\leq 0 
    \quad \quad \quad  {\rm for} \, \sigma_x =0
\end{align}
When no stress is applied, $L_x^{\rm SP}$ becomes shorter with time until the sample is fully equilibrated (i.e. self-healed).  The magnitude of the rate $\dot{L}_x^{\rm SP}$ during self-healing is usually much smaller than that during loading.

The three rules stated above provide some constraints and guidelines on the future development of a constitutive theory of DPNs based on the evolution of network microstructures.

\section{Conclusions} \label{sec:conclusions}
In summary, we constructed a CGMD model for elastomers with dynamic bonds (i.e. DPNs) that successfully captures the stress-strain hysteresis and self-healing behavior under uniaxial loading cycles observed in experiments. 
We present evidence that the average length of shortest paths between distant nodes in the network is a key microstructural parameter controlling the mechanical behavior of DPNs.
In addition, we show that the depiction of the DPN evolution as trajectories on the SP-strain map provides useful insights on the deformation mechanisms.
We present a set of empirical rules constraining the evolution of SP under various loading paths.
Our findings pave the way for the development of constitutive theories of DPNs that account for the evolution of network topology during deformation and self-healing.
Such physics-based constitutive theories could lead to new design rules for the next generation of DPNs.

\bigbreak

\section*{Methods} \label{sec:methods}
 We performed CGMD simulations using LAMMPS~\cite{plimpton1995fast} on a well-established bead-spring model~\cite{kremer1990dynamics} subjected to periodic boundary conditions. Each model contains 500 chains, each with 500 beads. Non-bonded interactions between each pair of beads were modeled by a Lennard-Jones (LJ) potential: $U_{\rm LJ}(r) = 4\varepsilon[(a/r)^{12}-(a/r)^6]$ for $r<r_c$, where $r$ is the inter-bead distance, $\varepsilon = 2.5$~kJ/mol~is the strength of the potential, $a = 15$\,\AA~is the diameter of the bead, and $r_c = 2.5a$ is the cutoff. The backbone interactions between neighboring beads on the same chain were modeled by a FENE potential: $U_{\rm FENE}(r) = -(kR_0^2/2)\ln[1-({r/R_0})^2]$,
with bond stiffness $k = 30\varepsilon/a^{2}$ and maximum bond length $R_0 = 1.5 a$. 
Before adding cross-links, we equilibrated the polymer melts for $23$~$\mu$s by using the NPT ensemble with a time step of 0.23 ps at room temperature and zero pressure~\cite{sliozberg2012fast,yin2020topological}. 

We use the LJ potential with modified parameters to model the dynamic cross-linking interaction between beads A and B because this functional form is computationally efficient and can reasonably capture the nature of the dynamic bonds in experiments. The potential parameters of the A-B interaction ($\varepsilon_{AB} = 15\varepsilon$ and $a_{AB} = a$) were set to be different from those of the regular LJ potential in a way that makes the strength of A-B potential comparable to that of the FENE potential. To avoid aggregation of multiple type-A beads around one type-B bead, the interactions of A-A and B-B were set to pure repulsion, which can be achieved by truncating the LJ potential at its minimum ($r_c = 2^{1/6}a$) with the following parameters: $\varepsilon_{AA} = \varepsilon_{BB} = 0.7\varepsilon$, $a_{BB} = 2.6a$ and $a_{AA} = 2.1a$. The parameters of A-A, B-B and A-B were switched gradually from the regular LJ parameters during a $460$-ns NPT simulation. After that, we ran a $138$-$\mu$s NPT simulation to achieve a network model cross-linked by approximately 8000 A-B-A bonds. The resulting structure measures 98.3~nm in all 3 dimensions~\cite{yin2020elastomers}.
For the loading and unloading simulations, we used the NVT ensemble (constant volume) at a strain rate of $8.7\times10^{5}\,{\rm s}^{-1}$. The second configuration was relaxed for $107.4$~$\mu$s with the NPT ensemble (to mimic the healing process in experiments) and then stretched to $400\%$ strain.




\section*{Conflict of Interest}
The authors declare no competing interests.
\section*{Author Contributions}
Y.Y., S.M., and W.C. designed the research approach. Y.Y., S.M., and W.C. performed the research. Y.Y. and S.M. analyzed the data, and Y.Y., S.M, C.C., Z.B., and W.C. wrote the paper.
\section*{Acknowledgements}
Y.Y., C.C., and Z.B. acknowledge support from Samsung Electronics. S.M. and W.C. acknowledge support from the Air Force Office of Scientific Research under award number FA9550-20-1-0397.

\section*{References}

\end{document}